\documentclass[twocolumn,preprintnumbers,showpacs,showkeys,amsmath,amssymb]{revtex4}

\usepackage{graphicx}

\begin{document}

\title{Electron Energy Spread in Laser Cooling Process }

\author{A. Kolchuzhkin}
\email{amk@chair12.phtd.tpu.edu.ru} 
\author{A. Potylitsyn}
\author{S. Strokov}

\affiliation{Tomsk Polytechnic University, Tomsk, Russia, 634034}

\begin{abstract}
The problem of electrons energy spread in the linear back Compton scattering 
process has been considered in the paper. The adjoint kinetic equation for the
electrons passing through a laser flash has been obtained and used to get the 
equations for the mean energy and the variance of the energy distribution.
The equations for the distribution moments have been obtained and solved by the 
iteration method in the continuous slowing down approximation with approximate 
description of the energy loss fluctuation. 
It has been shown that the variance of the energy distribution as a function 
of the  electron - photon collisions number in laser flash has a maximum for 
the electron beam with small incident
energy spread.  The beam energy spread damping in the laser-electron 
storage ring has been studied.
The data of approximate analytical calculations are compared with the results
of the Monte Carlo simulation.
\end{abstract}



%\end{frontmatter}

\maketitle

\section{Introduction}
The interaction of a high energy electron beam with an intense
laser flash is now considered in such projects as compact x-ray sources,
$\gamma -\gamma$ colliders, diagnostics of low emittance 
electron beams, laser cooling and others \cite{Angelo}. 
The laser flash intensity in these projects may be so high
that an electron can undergo several successive collisions passing through a 
photon bunch \cite{Telnov-83,Telnov-95}. It means the number of scattered
photons is larger than one. Taking into account the discrete character of the 
Compton back-scattering process, one can see an analogy between the electron
passage through a laser  flash and through a condensed
matter. The stochastic theory of particles penetration
through a matter can be found in the book \cite{We}. The
model of multiple bremsstrahlung process which leads to emission of a  few
photons by a single electron was developed in \cite{brems}.
Here we consider the multiple Compton back-scattering process using
the same approach. For simplicity we use the
approximation of an uniform photon concentration in the laser flash:
$$n_L = \frac{ A}{\omega_0} \ \frac{1}{\pi r^2_Ll_L}=
\rm const \,
$$
where $ A$ is the total energy of the laser flash; $\omega_0$ is the photon
energy, $r_L$ and $l_L$ are  radius and length of the photon bunch.
In other words our approximation corresponds to the short laser
pulse ($l_L \ll z_R$, $z_R$ being the Rayleigh length).

The kinetic equations have been obtained for monoenergetic electrons and
for electrons with incident energy spread.
These equations are transformed into the equations for the distributions 
moments. The method for an approximate solution of the equations  has been 
derived for the mean energy and for the variance of the energy distributions. It 
has been shown that the energy distribution variance as a function of 
the target thickness is a curve with a maximum if the incident energy spread is 
small. 

It is shown that the equation for calculation of the variance can be used 
to study the variation of the energy spread in the laser-electron storage 
ring (LESR) where an electron beam repeatedly encounters the laser pulses.
It has been shown that the variation of the energy spread in this case depends 
on the incident value of the variance. The energy spread damping takes place 
if the incident
 energy spread is larger than some limit value.

The results of analytical calculations are compared with the data of Monte Carlo 
simulation.

\section{Transport equations}
Let us consider electrons with an incident energy $\varepsilon_0\gg mc^2$ 
travelling 
through a bunch of photons with an energy $\omega_0$. The process of an electron 
transport in a light target is a random process where both the collisions 
number 
along the path $l$ and the energy loss in each collision are random quantities. 
Therefore, the energy of electron $\varepsilon$ passing a laser flash is 
random quantity too.

The angular deflection of an electron in the Compton back-scattering process is
small ($\displaystyle \sim\frac{2\omega_0}{mc^2}$) and may be neglected (the 
straight-ahead approximation). In this approximation the probability density 
function  $ P_m(\varepsilon|\varepsilon_0,l)$ (the subscript ''m'' means 
''monoenergetic'') describing the energy distribution of electrons 
travelling a path $l$ in the photon bunch obeys the 
adjoint balance equation (the Kolmogorov-Chapman equation) \cite{Feller,We}:
\begin{eqnarray}
&&P_m(\varepsilon|\varepsilon_0,l)=
(1-s\Sigma 
(\varepsilon_0))P_m(\varepsilon|\varepsilon_0,l-s)\nonumber \\
& &+s\Sigma (\varepsilon_0) \int\limits_{0}^{\omega_{max}}
\frac{\Sigma (\omega;\varepsilon_0) }{\Sigma (\varepsilon_0)}
P_m(\varepsilon|\varepsilon_0-\omega,l)d\omega\, ,
\label{sec:Peps}
\end{eqnarray}
where $\Sigma (\varepsilon_0)$ and $\Sigma (\omega;\varepsilon_0)$ are the total
and differential macroscopic cross-sections of the Compton scattering, 
$s$ is a small part of $l$, and $\omega$ is the energy of scattered
photon $\displaystyle (0\le \omega\le\omega_{max}),$
\[\omega_{max}=\varepsilon_0 \frac{x}{1+x}\]
is the maximum value of the scattered photon energy, 
\[\displaystyle x=\frac{4\omega_0 \varepsilon_0}{(mc^2)^2}\, ,\]
\[\Sigma(\varepsilon_0)=2n_{L}\sigma(\varepsilon_0)\, ,\]
\[\Sigma (\omega;\varepsilon_0)=
2n_{L}\frac{d\sigma(\omega;\varepsilon_0)}{d\omega}\, , \]
$\sigma$ and $\displaystyle \frac{d\sigma}{d\omega}$ are the total and 
differential cross-sections.
% and
 $n_{L}$ is the concentration of laser photons in the bunch.

Note that $\Sigma(\varepsilon_0)$ is the 
mean number of collisions of an electron per unit path length and 
$\Sigma (\omega;\varepsilon_0)$ is the mean number of an electron 
collisions with the energy loss in unit interval about $\omega$
per unit path length.

The first term in the right side of Eq. (\ref{sec:Peps})
corresponds to the electrons, which passes the path \( s \) without collisions, 
and $1-s\Sigma (\varepsilon_0)$ is corresponding probability. These 
electrons have to lose the energy $\varepsilon_0-\varepsilon$
along the rest path \( l-s \).

The second term corresponds 
to the electrons, which undergo the first scattering passing the path \( s \), 
and \( s\Sigma (\varepsilon_0) \) is the corresponding probability. The energy 
of the electron after the first scattering equals  $\varepsilon_0-\omega$, 
where \( \omega \) is a random energy of the scattered photon, and 
\( \Sigma (\omega;\varepsilon_0)/\Sigma (\varepsilon_0) \)
is the probability density function of $\omega$. In the limit 
\( s\rightarrow 0 \)
Eq. (\ref{sec:Peps}) gives the adjoint integro-differential equation for the
function
\( P_m(\varepsilon|\varepsilon_0,l) \) \cite{brems,stochastics}:
\begin{eqnarray}
&&\frac{\partial }{\partial l}P_m(\varepsilon|\varepsilon_0,l)+
\Sigma (\varepsilon_0)P_m(\varepsilon|\varepsilon_0,l)
\nonumber\\
& &-\int\limits^{\omega_{max}}_0
\Sigma (\omega;\varepsilon_0)
P_m(\varepsilon|\varepsilon_0 -\omega,l)d\omega=0
\label{PQ}
\end{eqnarray}
with boundary condition
\[P_m(\varepsilon|\varepsilon_0,l)|_{l=0}=
\delta (\varepsilon-\varepsilon_0),\]
$\delta(\varepsilon-\varepsilon_0)$ being the Dirac $\delta$-function. 

If the electrons incident on the light target have an energy distribution
$P_0(\varepsilon)$, the energy 
distribution of electrons behind the bunch is the 
convolution of $P_0(\varepsilon)$ and \( P_m(\varepsilon|\varepsilon_0,l) \): 
\begin{equation}
P_c(\varepsilon|l)=
\int\limits_{\varepsilon}^{\infty}
P_0(\varepsilon')P_m(\varepsilon|\varepsilon',l)d\varepsilon' 
\label{sec:Pnswoln}
\end{equation}
(the subscript ''c'' means ''convolution'').

Eq. (\ref{PQ}) can be transformed into the equation for the distribution 
moments 
\[\overline{\varepsilon_m^{k}}(\varepsilon_0,l)=
\int\limits^{\varepsilon_0}_{mc^2}
\varepsilon^{k}P_m(\varepsilon|\varepsilon_0,l)d\varepsilon.\]

The equation for $\overline{\varepsilon_m^{k}}(\varepsilon_0,l)$  
\cite{stochastics} is
\begin{eqnarray}
&&\frac{\partial }{\partial l}\overline{\varepsilon_m^{k}}(\varepsilon_0,l)+
\Sigma (\varepsilon_0)\overline{\varepsilon_m^{k}}(\varepsilon_0,l)
\nonumber\\
& &-\int\limits^{\omega_{max}}_{0}\Sigma (\omega;\varepsilon_0)
\overline{\varepsilon_m^{k}}(\varepsilon_0-\omega,l)d\omega=0.
\label{Q1}
\end{eqnarray}

The boundary condition for the moments is
$\overline{\varepsilon_m^{k}}(\varepsilon_0,l)$ is
\[\overline{\varepsilon_m^{k}}(\varepsilon_0,l)|_{l=0}=
\varepsilon_0^{k}.\]

The equation for the moments of the distribution $P_c(\varepsilon|l)$ describing 
transport of electrons with an incident energy spread can be 
obtained from Eq. (\ref{sec:Pnswoln}):
\begin{equation}
\overline{\varepsilon_c^{k}}(l)=
\int\limits_{mc^2}^\infty \varepsilon^{k} P_c(\varepsilon|l)d\varepsilon 
=\int\limits_{mc^2}^{\infty}
 P_0(\varepsilon')
\overline{\varepsilon_m^{k}}(\varepsilon',l) d\varepsilon'.
\label{sec:Mom-sig}
\end{equation}
It is seen that they are expressed in terms of the moments for the 
monoenergetic electrons and the incident energy distribution.

If the energy loss of an electron in one collision is small, the 
integro-differential equation for the moments (\ref{Q1}) can be
transformed by the Taylor expansion of integrands:

\begin{eqnarray*}
\overline{\varepsilon_m^{k}}(\varepsilon_0-\omega,l)&\approx&
\overline{\varepsilon_m^{k}}(\varepsilon_0,l)-
\omega\frac{\partial }{\partial\varepsilon_0}
\overline{\varepsilon_m^{k}}(\varepsilon_0,l)
\\
&&+\frac{1}{2}\omega^2\frac{\partial^2 }{\partial\varepsilon_0^2}
\overline{\varepsilon_m^{k}}(\varepsilon_0,l).
\end{eqnarray*}

This gives the partial differential equation:                                       
\begin{eqnarray}
\frac{\partial }{\partial l}\overline{\varepsilon_m^{k}}(\varepsilon_0,l)&+&
\beta (\varepsilon_0)
\frac{\partial }{\partial \varepsilon_0}
\overline{\varepsilon_m^{k}}(\varepsilon_0,l)
\nonumber\\
&-&\frac{1}{2}\gamma (\varepsilon_0)
\frac{\partial^2 }{\partial \varepsilon_0^2}
\overline{\varepsilon_m^{k}}(\varepsilon_0,l)=0.
\label{Q1csd}
\end{eqnarray}

The quantities  
\(\beta (\varepsilon_0) \), and 
\(\gamma (\varepsilon_0) \) in (\ref{Q1csd})
are the moments of the macroscopic differential cross-section:
\[
\beta (\varepsilon_0)=
\int\limits_{0}^{\omega_{max}}\omega\Sigma (\omega;\varepsilon_0)d\omega,
\]

\[
\gamma (\varepsilon_0)=
\int\limits_{0}^{\omega_{max}}\omega^{2}
\Sigma (\omega;\varepsilon_0)d\omega.
\]

Two first terms in Eq. (\ref{Q1csd}) correspond to the continuous slowing down 
approximation, whereas the third term gives approximate description of 
the energy loss fluctuations.  

The Eq. (\ref{Q1csd}) written for $k=1,~2$ can be 
transformed into the equation for the variance 
\[
\sigma_m^2(\varepsilon_0,l)=
\overline{\varepsilon_m^{2}}(\varepsilon_0,l)-
\overline\varepsilon_m^2(\varepsilon_0,l).\]
The equation has a form
 \cite{stochastics}
\begin{eqnarray}
\frac{\partial }{\partial l}\sigma_m^2(\varepsilon_0,l)+
\beta (\varepsilon_0)
\frac{\partial }{\partial \varepsilon_0}
\sigma_m^2(\varepsilon_0,l)
-\frac{1}{2}\gamma (\varepsilon_0)
\frac{\partial^2 }{\partial \varepsilon_0^2}
\sigma_m^2(\varepsilon_0,l)\nonumber\\
=\gamma (\varepsilon_0)\left(\frac{\partial }{\partial \varepsilon_0}
\overline{\varepsilon_m}(\varepsilon_0,l)\right)^2.
\label{Delta}
\end{eqnarray}

The Eq. (\ref{sec:Mom-sig}) for the energy distribution moments in the case of 
the electron beam with an incident energy spread can be simplified if the 
distribution $P_0(\varepsilon)$ 
is a symmetric function with maximum at a point $\varepsilon_0$ and small 
variance 
\[\sigma_0^2=\int\limits_{-\infty}^{\infty}
(\varepsilon-\varepsilon_0)^2~P_0(\varepsilon)d\varepsilon.\]
In this case the Taylor expansion of the function 
$\displaystyle\overline{\varepsilon_m^{k}}(\varepsilon,l)$ in 
(\ref{sec:Mom-sig}) makes it possible to express the moments of the distribution
$P_c(\varepsilon|l)$ in terms of the moments 
$\overline{\varepsilon_m^{k}}(\varepsilon_0,l)$ corresponding to 
monoenergetic electrons:
\begin{equation}
\overline{\varepsilon_c^{k}}(l)=
\overline{\varepsilon_m^{k}}(\varepsilon_0,l)
+\frac{\sigma_0^2}{2}
\frac{\partial^2 }{\partial \varepsilon_0^2}
\overline{\varepsilon_m^{k}}(\varepsilon_0,l).
\label{sec:mean-d}
\end{equation}
The equations (\ref{sec:mean-d}) written for $k=1,2$ can be easily transformed 
into the equation for the variance 
\[\sigma_c^2(l)=\overline{\varepsilon_c^2}(l)-\overline{\varepsilon_c}^2(l).\]
The equation is
\begin{eqnarray}
\sigma_c^2(l)&=&\sigma_m^2(\varepsilon_0,l)\nonumber\\
&&+\frac{\sigma_0^2}{2} 
\left(\frac{\partial^2 }{\partial \varepsilon_0^2}\sigma_m^2(\varepsilon_0,l)
+2( \frac{\partial}{\partial \varepsilon_0}
\overline{\varepsilon_m}(\varepsilon_0,l))^2\right).
\label{sec:sigma-d}
\end{eqnarray}
It describes the variance of the energy distribution for the electron beam
with an incident variance $\sigma_0$ in 
terms of stochastic characteristics for monoenergetic electrons.

\section{Compton scattering cross-sections and interaction coefficients}

In this paper we restrict our consideration to the linear Compton scattering. 
The differential cross-section of this process for relativistic electron
is
\begin{eqnarray}
&&\frac{d\sigma(y;x)}{dy}=\frac{2\pi r_0^2}{x} \nonumber\\
&&\times\left(1-y+\frac{1}{1-y}
-\frac{4y}{x(1-y)}+\frac{4y^2}{x^2(1-y)^2} \right) ,
\label{compton}
\end{eqnarray}

where $\displaystyle y=\frac{\omega}{\epsilon_0}$ and
$\displaystyle r_0=\frac{e^2}{mc^2}$ is the classical radius of 
electron \cite{Telnov-83}. 

In the energy region of our interest the invariant dimensionless parameter 
$x$ is small $(x\ll 1)$ and the interaction coefficients
\(\beta (\varepsilon_0) \), and 
\(\gamma (\varepsilon_0) \) 
are described by the approximate formulas  
\begin{equation}
\label{beta}
\beta (\varepsilon_0)
\approx \frac{\Sigma_0}{2}\varepsilon_0 x\, 
\end{equation}
\begin{equation}
\label{gamma}
\gamma (\varepsilon_0)
\approx \frac{7}{20}\Sigma_0 \varepsilon_0^2 x^2, 
\end{equation}

where $\displaystyle \Sigma_0=2 n_L \sigma_T$, 
$\displaystyle \sigma_T=\frac{8}{3}\pi r_0^2$
being the Thomson cross-section.

The quantities $\beta(\varepsilon_0)$ and $\gamma(\varepsilon_0)$ are the mean 
energy loss 
and the mean squared energy loss of an electron per unit path length.

\section{Solution of equations for monoenergetic electrons}

The partial differential equations (\ref{Q1csd}), (\ref{Delta}) with 
the interaction coefficients (\ref{beta}), (\ref{gamma}) describe 
the mean energy and the variance of the energy distribution for incidentally 
monoenergetic electrons passing a path $l$. The equations can be 
solved by the iteration method. In the first approximation, where the terms 
with the second derivatives are neglected, the solutions are

\begin{equation}
\overline{\varepsilon_m }(\varepsilon_0,l)=
\frac{\varepsilon_0 }{1+\frac{1}{2}\overline{n} x}\, 
\label{sec:eps}
\end{equation}
\begin{equation}
\sigma_m^2(\varepsilon_0,l)=
\frac{7}{20}\frac{\varepsilon_0^2 x^2 \overline n}{\left( 1+
\frac{1}{2} \overline{n} x\right)^4},
\label{sec:var}
\end{equation}
where $\overline n=\Sigma_0 l$ is the mean number of collisions in the 
light target \cite{stochastics}.

The mean value (\ref{sec:eps}) is a monotonically decreasing function of variable 
$l$. 
The Taylor expansions show that for small $l$ $(\overline n x\ll 1)$
\[\overline{\varepsilon_m }(\varepsilon_0,l)\approx \varepsilon_0 (1-
\frac{\overline n x}{2})\]
and for large $l$ $(\overline n x\gg 1)$
\[\overline{\varepsilon_m }(\varepsilon_0,l)\approx 
\frac{2\varepsilon_0}{\overline n x}\left(1-\frac{2}{\overline n x}\right).\]

Similarly, it can be shown  that the variance (\ref{sec:var}) grows with 
$l$ for thin targets:
\[\sigma_m^2(\varepsilon_0,l)\approx\frac{7}{20}\overline n \varepsilon_0^2 x^2
(1-2 \overline n x)\]
and decreases for large $l$:
\[\sigma_m^2(\varepsilon_0,l)
\approx\frac{28}{5}\frac{\varepsilon_0^2}{\overline n^3 x^2}
(1-\frac{8}{\overline n x}).\]

It follows from (\ref{sec:var}) that the variance $\sigma_m^2(\varepsilon_0,l)$ 
has a maximum at the point, where 
\[\displaystyle \overline n=\frac{2}{3}\frac{1}{x}\]
or
\[\displaystyle \frac{\overline \varepsilon_m(\varepsilon_0,l)}
{\varepsilon_0}=\frac{3}{4}.\]

The energy spread damping for large $l$
is due to the energy loss $\beta(\varepsilon)$ per unit path 
length is 
proportional to $\varepsilon^2$ ,therefore, higher energy electrons in a 
bunch lose more energy than lower energy electrons \cite{Telnov-97,Huang1}.

Using Eqs. (\ref{sec:var}) and (\ref{sec:eps}) 
one can derive the formula 
\[\sigma_m^2(\varepsilon_0,l)=
\frac{7}{10}x\frac{\overline \varepsilon}{\varepsilon_0}
\left(1-\frac{\overline \varepsilon}{\varepsilon_0}\right),\]
which was earlier obtained in \cite{Telnov-97}. 

\section{Solution of equations for electron beam with incident energy spread}

Substitution of Eqs. (\ref{sec:eps}), (\ref{sec:var}) in Eqs. 
(\ref{sec:mean-d}) and (\ref{sec:sigma-d}) makes it possible to calculate 
the mean energy and the variance of the energy distribution for the beam with 
the incident energy spread $\sigma_0$.

These calculations show that the mean value $\overline{\varepsilon_c}(l)$ is 
decreased monotonically with $l$:
\[\overline{\varepsilon_c }(l)\approx \varepsilon_0\left(1-
 \frac{\overline n x}{2}(1+\frac{\sigma_0^2}{\varepsilon_0^2})\right)\]
for small $l$ and
\[\overline{\varepsilon_c }(l)\approx 
\frac{2\varepsilon_0}{\overline n x}
\left(1-
\frac{2}{\overline{n} x}(1+
\frac{\sigma_0^2}{\varepsilon_0^2})\right)\]
for large $l$.

But the dependence of variance $\sigma_c^2(l)$ on $l$  is determined by the
incident energy spread. 
The Taylor expansion of Eq. (\ref{sec:sigma-d}) in powers of $l$ gives
\[\sigma_c^2(\varepsilon_0,l)\approx
\sigma_0^2 + \frac{7}{20}\varepsilon_0^2 \overline n x^2
\left(1-\frac{40 \sigma_0^2}{7 \varepsilon_0^2 x}(1-\frac{21 x}{20})\right),\]
whereas for large $l$ it is decreased with $l$:
\[\sigma_c^2(\varepsilon_0,l)
\approx\frac{28}{5}\frac{\varepsilon_0^2}{\overline n^3 x^2}\left(1
-\frac{8}{ \overline{n} x}\left(1+\frac{\sigma_0^2}{\varepsilon_0^2}
(1-\frac{10}{7 x})\right)\right).\]
It means that the variance $\sigma_c^2(l)$ has a maximum at some $l$ if 
$\sigma_0^2\le \sigma^2_{max}$, where
\begin{equation}
\sigma^2_{max}=\frac{7}{40} x \varepsilon_0^2 \frac{1}{1-\frac{21}{20}x},
\label{sec:sig-max}
\end{equation}
 
For $\sigma_0= \sigma_{max}$ this maximum is at the point $l=0$.
If 
 $\sigma_0 > \sigma_{max}$ the variance $\sigma_c^2(l)$ is a decreasing 
function of $l$.

\section{Multiple crossing of laser bunch}

It was pointed out in \cite{Huang1,Huang2} that effective radiative cooling 
of an electron beam can be realized in LESR where electrons repeatedly interact 
with an intense laser pulses. Eq. (\ref{sec:sigma-d}) can be used to study the 
variation of the energy spread for electrons multiply crossing
the light target.  If the lost energy is restored by an rf accelerating system 
after each turn, the variance of energy distribution after n-th
turn can be calculated using Eq. (\ref{sec:sigma-d}) with $\sigma_0^2$ replaced by 
the variance after (n-1)-th turn:
\begin{equation}
\label{iterat}
\sigma^2(n)=\sigma_m^2(\varepsilon_0,l)+
\eta~ \sigma^2(n-1),
\end{equation}
where
\begin{equation}
\label{eta}
\eta=\frac{1}{2}
\left(\frac{\partial^2 }{\partial \varepsilon_0^2}\sigma_m^2(\varepsilon_0,l)
+2( \frac{\partial}{\partial \varepsilon_0}
\overline{\varepsilon_m}(\varepsilon_0,l))^2\right).
\end{equation}

It follows from (\ref{iterat}) that for $|\eta|\le 1$
\begin{equation}
\sigma^2(\infty)
=\sigma_m^2(\varepsilon_0,l)(1+\eta+\eta^2+\eta^3+...)=
\frac{\sigma_m^2(\varepsilon_0,l)}{1-\eta}. 
\label{sec:var-inf}
\end{equation}

Let us point out that the multiple crossing of the laser bunch leads to the 
energy spread damping if the incident energy spread 
$\sigma_0^2>\sigma^2(\infty)$. In the opposite case 
($\sigma_0^2<\sigma^2(\infty)$) the energy spread is increased.

Limit value of energy spread $\sigma^2(\infty)$ depends on the thickness of
laser bunch especially for high energy electrons. For small $l$
\[\sigma^2(\infty)\approx
\sigma^2_{max}\left(1-\overline n~x~\frac{(15+28 x)}
{(20-21 x)}\right)
,\]
whereas for large $l$ 
\[\displaystyle\sigma^2(\infty)\to\frac{28}{5}
\frac{\varepsilon_0^2}{\overline n^3 x^2}.\]

It is easy to see that for small $x$ 
\[\frac{\sigma(\infty)}{\varepsilon_0}\approx \frac{\sigma^2_{max}}{\varepsilon_0}
\approx \sqrt{\frac{7}{40}x}=
\sqrt{\frac{7}{10} \gamma \frac{\lambda_e}{\lambda_L}},\]
where $\lambda_L$ is the laser radiation wavelength and 
$\displaystyle \lambda_e=\frac{h}{m c}$ is the Compton wavelength of electron.
This result coincides with that one given in \cite{Huang2}. But our
approach allows to estimate the number of turns in a LESR to reach the limit 
value of the energy spread.

\section{Numerical results}

\begin{figure*}
\centering
\includegraphics{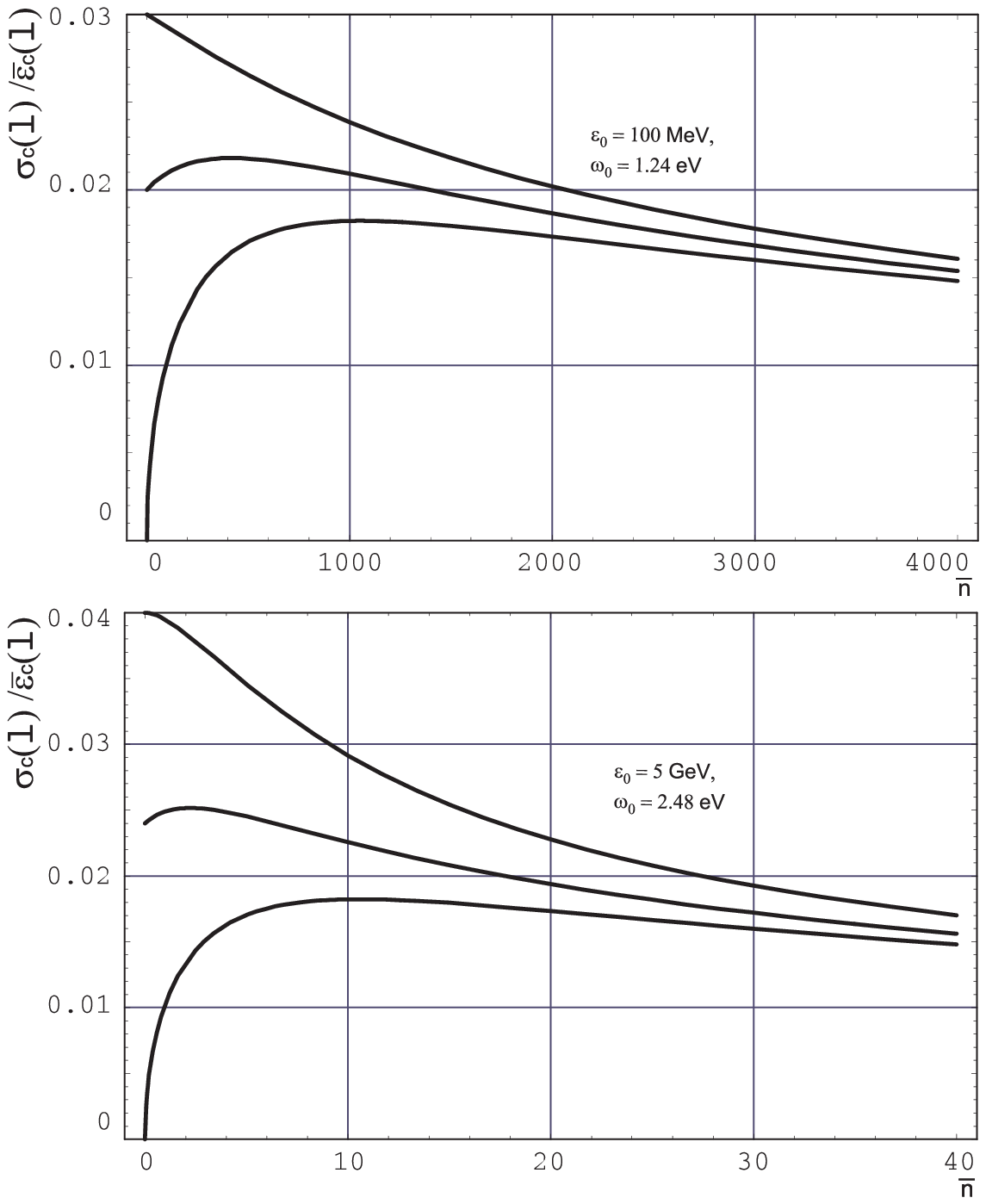}
\caption{The relative energy spread
$\displaystyle \sigma_c(l)/\overline{\varepsilon_c}(l)$
as a function of the target thikness for various initial energy spread.
}
\label{fig:fig1}
\end{figure*}

Numerical calculation of the mean energy and the variance of the energy 
distribution were made for $100~MeV$ electrons interacting with $1.24~eV$ 
laser photons and for $5~GeV$ electrons interacting with $2.48~eV$ photons.
Eqs. (\ref{sec:mean-d}) and (\ref{sec:sigma-d}) were used to calculate the 
relative energy spread
$\displaystyle \frac{\sigma_c(l)}{\overline{\varepsilon_c}(l)}$. The results are 
given in Fig. \ref{fig:fig1} for 3 values of the incident energy spread. 
They agree with the data obtained using Eq. (15) of \cite{Telnov3}.

%\begin{figure*}
%\centering
%\includegraphics{fig1_1.eps}
%\end{figure*}

It is seen from the figure that the relative energy spread of 
the electron beam as a 
function of the collisions number $\bar n $ is a decreasing function 
if the incident energy spread is large enough. But for the beams with 
small incident spread the ''heating'' of the beam takes place instead 
of ''cooling'' for small $l$. 

\begin{figure*}
\centering
\includegraphics{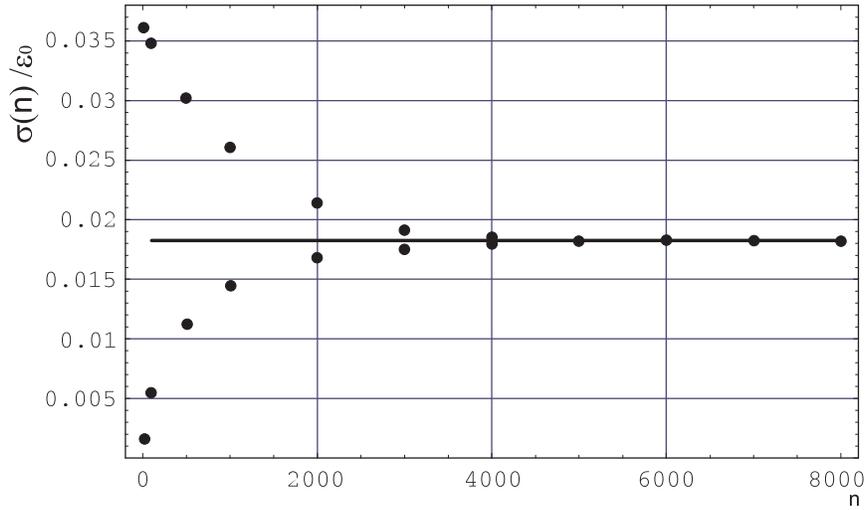}
\caption{The relative energy spread
as a function of the number of turns in LESR, 
$\varepsilon_0=100~MeV$, $\omega_0=1.24~eV$, and $\overline n=0.27$ \cite{Huang2}.
}
\label{fig:fig2}
\end{figure*}

The dependence of the energy distribution variance on the number of turns in 
LESR is shown in Fig. \ref{fig:fig2} for two 
incident energy spreads. The calculations were made for LESR  suggested for 
laser beam cooling in \cite{Huang2}. In this case $\omega_0=1.24~eV,~ 
\varepsilon_0=100~MeV$, the mean energy of scattered photon equals 
$\displaystyle \overline \omega =\frac{1}{2}\varepsilon_0 x=95~keV$, 
the average ring radius equals $1~m$, and the
energy loss per turn $\Delta \omega = 25~keV$. The mean number of 
collisions in laser bunch $\overline n=\frac{\Delta \omega}{\overline{\omega}}=0.27$.
It is seen from the figure that the asymptotic value of the energy spread is 
established in $\sim 5\cdot 10^3$ turns and it takes $\sim 100~\mu sec.$

It is seen from Fig. \ref{fig:fig2} that the energy spread 
of electrons multiply crossing
the light target is decreased or increased depending on its incident value.

\begin{figure*}
\centering
\includegraphics{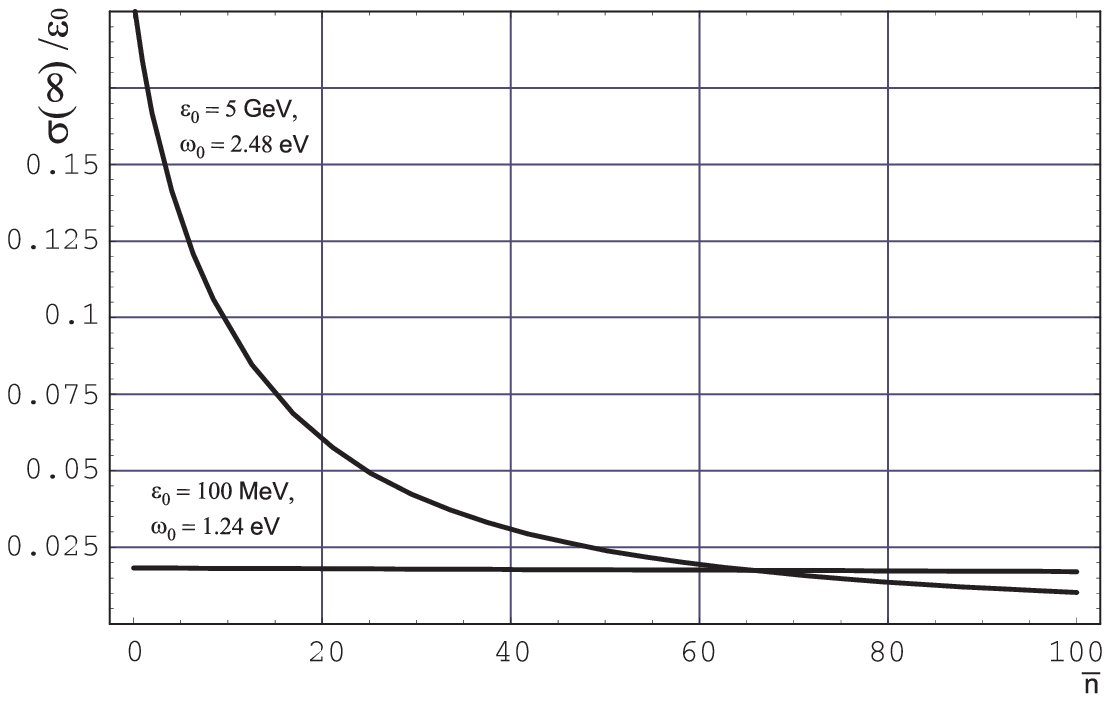}
\caption{
The limit value of the relative energy spread 
$\sigma(\infty)/\varepsilon_0$ as a function of the mean collisions number.
}
\label{fig:fig3}
\end{figure*}

The limit value of the relative energy spread 
$\sigma(\infty)/\varepsilon_0$ as a function of the light target thickness is 
given in Fig. \ref{fig:fig3}. It is seen from the figure that the variance
of the energy distribution is decreased with the target thickness increasing.

\section{Monte Carlo simulation}
It is supposed in our Monte Carlo simulation of the back Compton scattering that
the incident energy distribution of electrons is Gaussian with given variance
$\sigma^2_0$. The number of electron collisions with laser photons is supposed 
to be random. It is 
selected from the Poisson distribution with fixed \( \overline n \). The 
simulation of individual collisions is carried out in the electron rest frame 
using the Klein and Nishina formula with the Lorentz
transformation to the lab system. In the same way as in analytical 
calculations above we neglected the angular deflection of electrons but 
accounted for the energy decreasing after each collision.
% The energy 
%distributions of electrons are calculated and used to estimate the mean energy 
%and the variance of the distribution. The results are given in the figures 
%along with corresponding analytical data. 

\begin{figure*}
\centering
\includegraphics{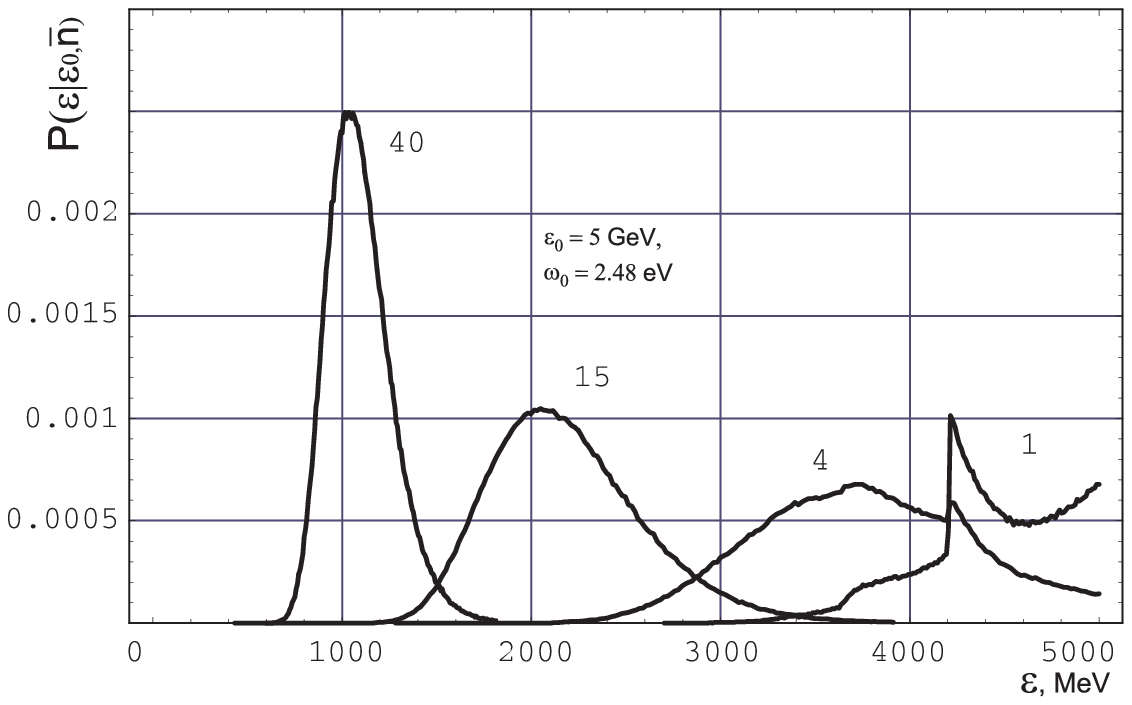}
\caption{The energy distributions of initially monoenergetic electrons 
for $\overline n$=1, 4, 15 and 40.
}
\label{fig:fig4}
\end{figure*}

Fig. \ref{fig:fig4} shows the energy distributions of $5~GeV$ electrons 
for several values of the light target thickness. It should be pointed out the 
discontinuous of the spectra for small $\overline n$ due to the single 
scattered electrons. It is seen from the figure that the energy spread 
increases with the target thickness increasing for small $\overline n$ and 
decreases
for such $\overline n$, which are greater than $\sigma^2_{max}$ determined by    
Eq. (\ref{sec:sig-max}). 

\begin{figure*}
\centering
\includegraphics{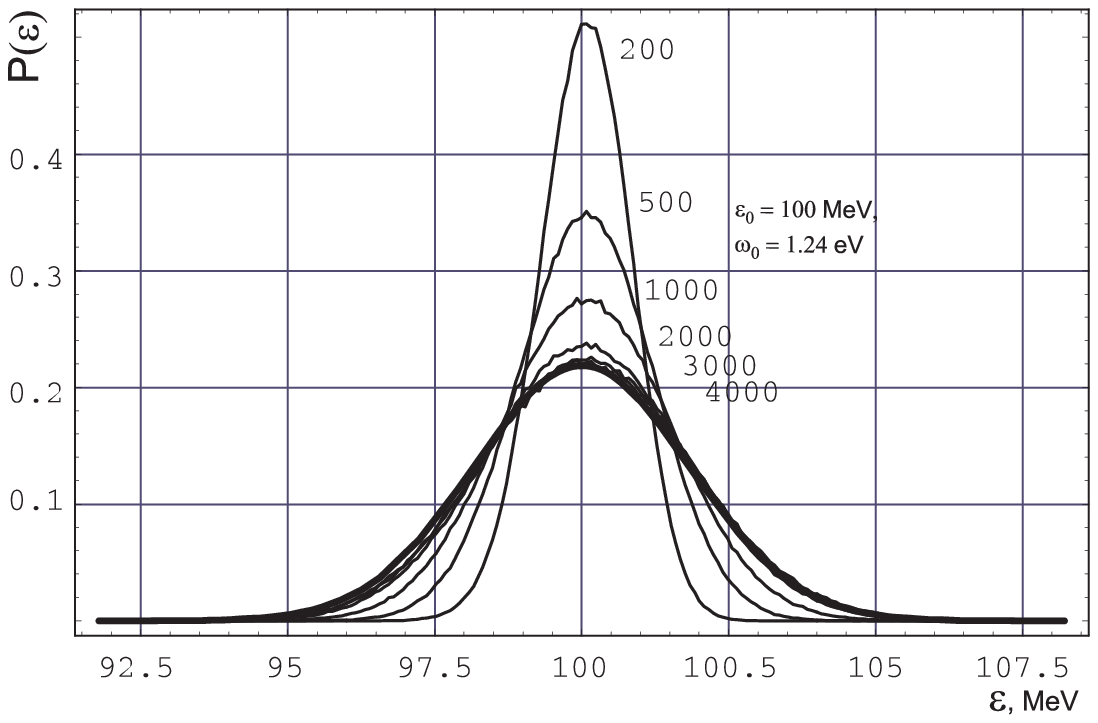}
\caption{Evolution of the energy distributions $P(\varepsilon)$ for 
electrons with $\sigma_0<\sigma(\infty)$ multiply crossing
the light target. Solid line is the Gaussian distribution with the
variance $\sigma(\infty)$. Number of turns is shown near the curves.}
\label{fig:fig5}
\end{figure*}

\begin{figure*}
\centering
\includegraphics{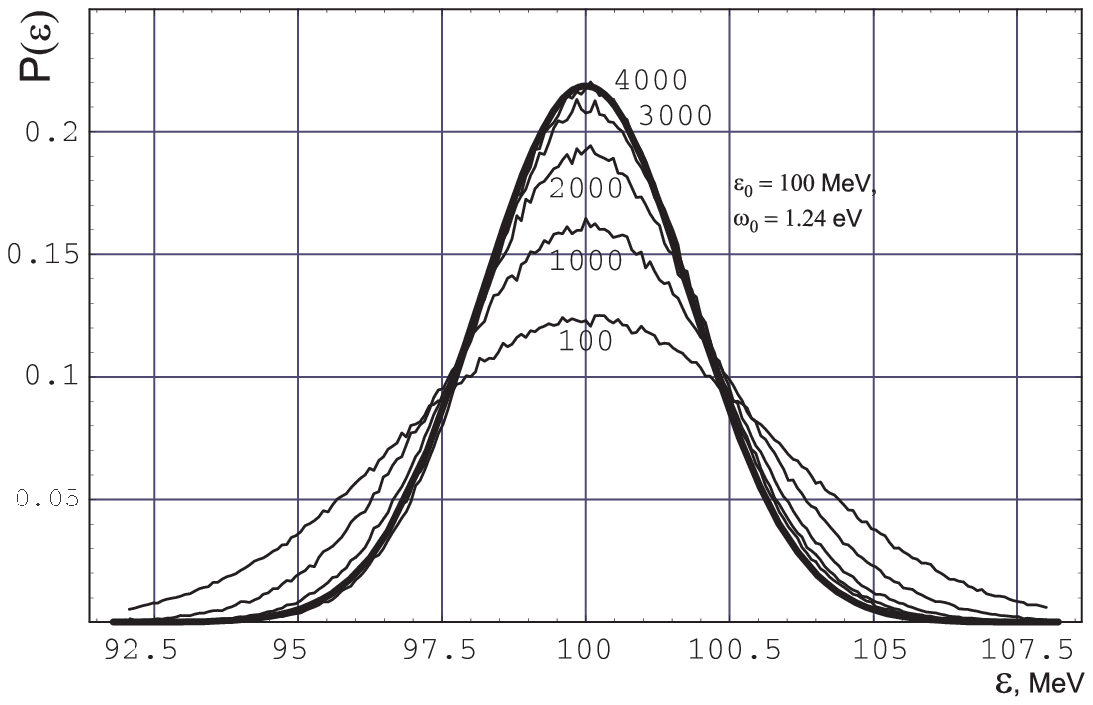}
\caption{Evolution of the energy distributions $P(\varepsilon)$ for 
electrons with $\sigma_0>\sigma(\infty)$ multiply crossing
the light target. Solid line is the Gaussian distribution with the
variance $\sigma(\infty)$. Number of turns is shown near the curves.}
\label{fig:fig6}
\end{figure*}

Fig. \ref{fig:fig5} and \ref{fig:fig6} shows the electron energy distributions in 
LESR for several number of turns.

All results are obtained with statistics more than $10^6$ trajectories.

It should be pointed out that the analytical formulas obtained in the continuous 
slowing down approximation with approximate consideration of the energy loss 
fluctuations are inapplicable for highly relativistic electrons 
if the mean energy loss in one collision is comparable to 
$\varepsilon_0$.  The Monte Carlo method has to be used in this case.  

\begin{figure*}
\centering
\includegraphics{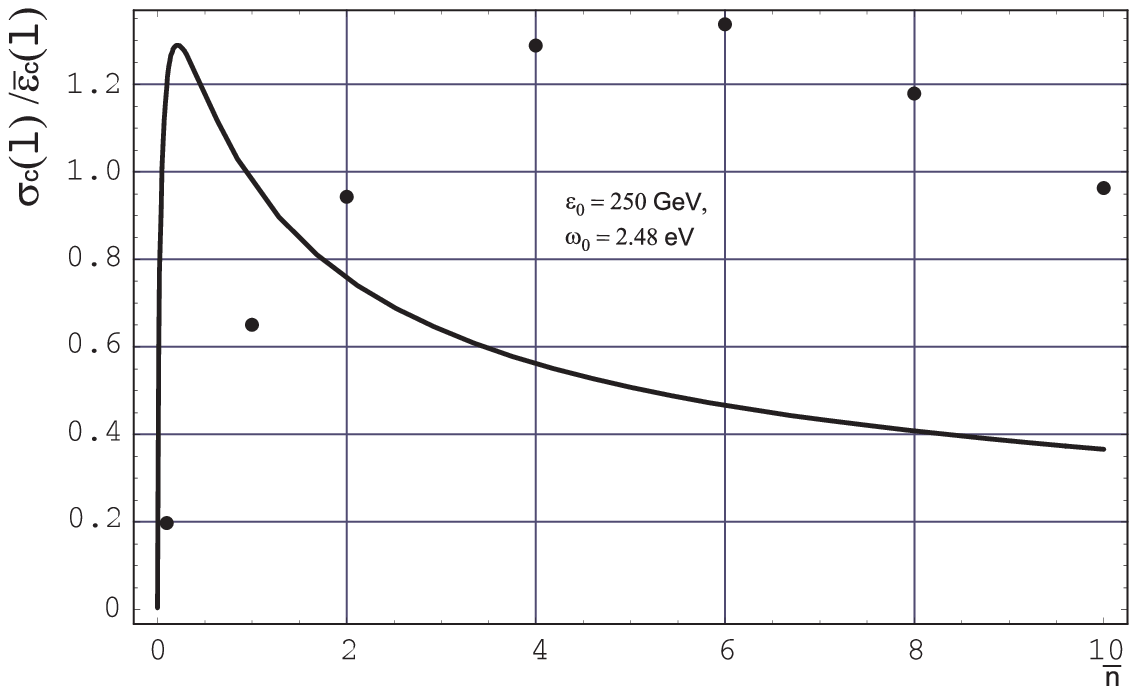}
\caption{The relative energy spread
$\displaystyle \sigma_c(l)/\overline{\varepsilon_c}(l)$
as a function of the target thickness.
 Solid line - Eqs. (\ref{sec:mean-d}), (\ref{sec:sigma-d}), 
points - Monte Carlo simulation.
}
\label{fig:fig7}
\end{figure*}

\begin{figure*}
\centering
\includegraphics{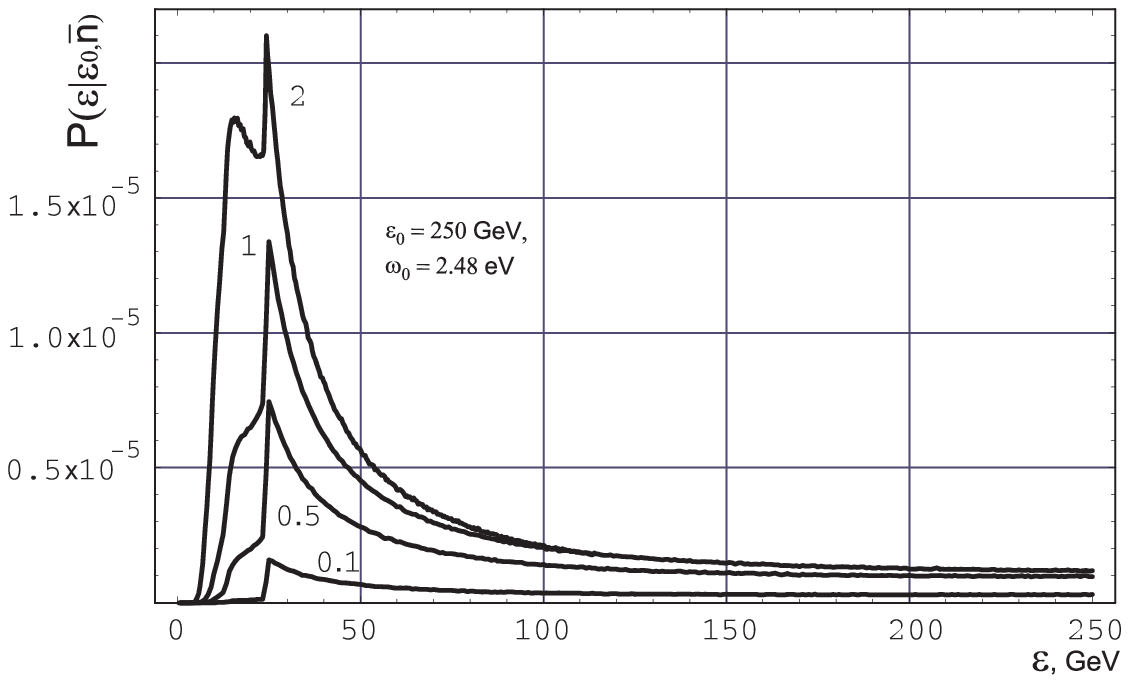}
\caption{The energy distributions of initially monoenergetic electrons 
for $\overline n$ = 0.1, 0.5, 1 and 2.
}
\label{fig:fig8}
\end{figure*}

Fig. \ref{fig:fig7} shows the relative energy spread as a function of the light target 
thickness calculated in continuous slowing down approximation and by the Monte
Carlo method for $250~GeV$ electrons. Fig. \ref{fig:fig8} shows the energy spectra of
these electrons.
Notice that the Monte Carlo technique allows to get the results with taking into
account the energy dependence of the interaction cross-sections and the lateral
distribution of photons in the laser flash.

\end{document}